# Effect of Imbalanced Datasets on Security of Industrial IoT Using Machine Learning

Maede Zolanvari
*Computer Science and Engineering*
*Washington University in St. Louis*
United States
maede.zolanvari@wustl.edu

Marcio A. Teixeira
*Department of Informatics*
*Federal Institute of São Paulo*
Brazil
marcio.andrey@ifsp.edu.br

Raj Jain
*Computer Science and Engineering*
*Washington University in St. Louis*
United States
jain@wustl.edu

***Abstract*—Machine learning algorithms have been shown to be suitable for securing platforms for IT systems. However, due to the fundamental differences between the industrial internet of things (IIoT) and regular IT networks, a special performance review needs to be considered. The vulnerabilities and security requirements of IIoT systems demand different considerations. In this paper, we study the reasons why machine learning must be integrated into the security mechanisms of the IIoT, and where it currently falls short in having a satisfactory performance. The challenges and real-world considerations associated with this matter are studied in our experimental design. We use an IIoT testbed resembling a real industrial plant to show our proof of concept.**

## 1 Introduction

Leveraging the internet of things (IoT) technology in the industrial control systems (ICSs), known as the industrial internet of things (IIoT), has become very popular in recent years. ICSs are the essential part of every critical infrastructure and have been utilized for a long time to supervise industrial machines and processes. Supervisory Control and Data Acquisition (SCADA) systems often manage the ICSs and are considered as the largest subset of these systems. Main roles of these systems are to perform real-time monitoring and interacting with the devices, real-time gathering and analyzing the data, and logging all the events that happen in the system. Utilizing IoT technology in these systems enhances the network intelligence and security in optimization and automation of industrial processes. IIoTs are mostly mission-critical applications with high-availability requirements. Their operations lead to a huge amount of data that can be easily managed through big data analysis methods.

In the past, to secure ICSs from malicious outside attack, these systems used to be isolated from the outside world. However, recent advances, increased connectivity with corporate networks, and utilization of internet communications to transmit the information more conveniently have introduced the possibility of cyber-attacks against these systems. Due to the sensitive nature of the industrial application, security is the foremost concern.

Since intrusion is the primary security concern in IIoT, an intrusion detection system (IDS) is an integral part of these applications to provide a secure environment. Stuxnet worm, which was exposed in 2010 [1] and recently reappeared (late December 2017), and Triton malware against the ICSs [2] raised awareness of the necessity for special attention to the security of these critical infrastructures. Through the fundamental differences between the ICSs and the regular IT systems, their common vulnerabilities and priorities are different [3]. Furthermore, ICSs have a specific type of traffic and data using particular IIoT communication protocols (e.g., Modbus, BACnet, DNP3). Due to all these reasons, proper diligence must be considered when it comes to designing an IDS for ICSs.

Machine learning-based security solutions have been widely used in providing security for IT systems. However, the suitability of these techniques for IIoT applications is debatable. The main security concern in IIoT devices is to detect any penetration into the system. Intrusion detection comes with special features such as significant imbalanced datasets that sometimes the trained machine learning (ML) algorithms may not be able to detect the attack.

In our previous works [4] and [5], we have designed different ML-based IDSs for ICSs through different attack scenarios, such as denial of service (DoS), SQL injection, and reconnaissance. However, we never truly studied the imbalanced datasets problem facing ML algorithms, where the real barriers are, and how different performance metrics would react to this problem. In this paper, after discussing how ML can be beneficial in IDS applications, we will study the cases where current machine learning algorithms fall short of providing the required level of security. More specifically, our main focus is on the imbalanced dataset problem in IIoT. The metrics that can fairly judge the performance have been compared to measure their effectiveness.

## 2 Related Work

In this section, we review some of the related research works. To the best of our knowledge, imbalanced IIoT dataset problem with the significantly low number of minority samples has not been studied yet.

The intrusion detection problem in smart grids using several different ML techniques has been studied in [6]. Some countermeasures to overcome the problem of imbalanced dataset have been examined. They have used ADFA-LD dataset that consists of 12.5% attack data. It is important to notice that this ratio is not realistic in the case of IIoT applications. Here we deal with less than 1% anomaly samples in our applications, which makes the results closer to real-world scenarios.

Various sampling techniques to overcome the imbalanced dataset problem have been investigated in [7]. The utilized datasets are extracted from Github and Sourceforge projects with 15% imbalance ratio. This work is not cybersecurity nor IoT related, though it represents a practical case study on this imbalanced dataset problem.

An IDS using a combination of J48 and Naive Bayes techniques is designed in [8]. The dataset was built using gas pipeline system of the Distributed Analytics and Security Institute, Mississippi State University, Starkville, MS. Their dataset consisted of different types of attacks such as reconnaissance, code injection, response injection, command injection. The J48 classifier was first used as a supervised attribute filter. Then, the Naive Bayes classifier was used to develop the anomaly-based intrusion detection. The ratio of attack traffic in their study was about 21.87%, which is far higher than a real-world case.

Six different types of ML algorithms, Naive Bayes, Random Forests, OneR, J48, NNge (non-nested generalized exemplars), SVM (support vector machines) for IDS have been studied in [9]. J48 is a type of decision tree technique. Their dataset consists of labeled RTU telemetry data from a gas pipeline system in Mississippi State University's Critical Infrastructure Protection Center. The attack traffic is generated from two types of code injection set, command injection attacks, data injection attacks. Seven different variants of data injection attacks were tried to change the pipeline pressure values, and four different variants of command injection attacks to manipulate the commands that control the gas pipeline. They used precision and recall metrics to make sure to have a fair evaluation in spite of the imbalanced dataset with about 17% attack traffic.

K-means technique, which is an unsupervised clustering algorithm, for IDS has been employed in [10]. An open-source virtual PLC (OpenPLC platform) along with AES-256 encryption is used to simulate an ICS. They have conducted three different types of attacks against their system, code injection, DoS, and interception (eavesdrop). However, they have not provided any information on the percentage of attack data that was used for training.

One class SVM (OCSVM) as a proper anomaly-based IDS has been proposed in [11]. They declare that OCSVM is a good choice because the dataset is imbalanced. The authors just used two features of traffic (data rate and packet size) of an electric grid. The trained model did not include any malicious attack data, and the trained dataset was captured during normal operation of a SCADA system.

## 3 WHY MACHINE LEARNING

IDS as an effective mechanism to counter intrusions has been widely used to provide a secure platform. Rule-based, signature-based, flow-based, and traffic-based are just some examples of different ways that intrusion detection has been implemented. Regarding the IIoT system, traditionally most of the connections and traffics in an ICS network were pre-defined. Hence, these types of IDS (ruled-based, signature-based, etc.) would detect abnormal activities very efficiently. For instance, when the intruder had to somehow manipulate the structure, like building new connections to the victims or sending a different type of traffic, ruled-based IDS would be successful in detecting the malicious attempt [12].

However, considering frequent upgrades in the networks, which results in regular changes in the topology, the legacy types of IDS will not work. Since these IDSs are designed based on defined topologies (e.g., allowed connections, allowed devices, etc.) any small changes in the system will raise a false alarm, unless the whole IDS would be re-designed after each change, which is an intensive task and might not work properly.

On the other hand, the legacy IDSs cannot keep up with the attackers constantly evolving their methods. Furthermore, to counter new attacks that appear every day, or in scenarios where the attack is planned perceptively (e.g., the man-in-the-middle attack), intelligent IDSs are required. An anomaly-based IDS that employs ML algorithms can detect any out of the ordinary activity if the training procedures are handled correctly.

The intelligent IDS is based on the fact that AI algorithms can detect anomaly patterns that are difficult for a human to discover. Unlike rule-based IDSs, ML-based IDSs can successfully detect new types of attacks, different variants of a specific attack and unknown or zero-day attack. The zero-day exploit takes advantage of unknown vulnerabilities (i.e., the developers have no idea that they exist) to manipulate the processes or the system. These are all the reasons that ML should be applied in designing IDSs.

## 4 WHERE MACHINE LEARNING FALLS SHORT

Despite the confidence in the ability of machine learning (ML) to detect anomalies very effectively, there exist several challenges that arise when considering their applicability in IIoT. Without addressing these problems, the ML algorithms are unable to function properly. In an IIoT environment, some of these challenges might be manageable and some might not, due to the nature of these systems and their associated security aspects.

The very first consideration is to choose proper features from the network traffic dataset. Sensor data in IIoT are usually obtained during an extended period from many sensors with different sampling frequencies, which results in high-dimensional datasets. Using raw data like this will add a large delay in training and detecting process. On the other hand, If the selected features do not vary during the attacks, even the best algorithm will not be able to detect an intrusion or an anomalous situation using that feature. It is, therefore, necessary to extract discriminating features to be able to use ML techniques. Applying power spectral density, Fourier analyses, the linear feature extracting method, and principal component analysis (PCA) are some examples of methods that could be tried out to reduce the dimensionality and find the most useful features.

On the other hand, due to the confidentiality and user privacy restrictions, industrial companies hardly release their protected network data on the intrusion attacks that might have occurred. Hence, training the ML algorithms on data collected from real networks of real industrial IoT's is almost impossible. Hence, most of the available research work in this area is done on commercial or public datasets that may not be specific to IIoT. That is another barrier in utilizing these ML techniques directly in companies or industrial networks. Due to this reason, we built our IIoT testbed and took all consideration into account to make it as resembling as possible to a real industrial plant. More details are provided in Section V.

Furthermore, in any real world IIoT system, the number of intrusion attack samples is significantly low. Since the intruders do not wish to be exposed; they usually run their attacks randomly in short periods of time. This leads to a very low amount of attack data to train the ML algorithm. This problem is known as an imbalanced training dataset. In other words, imbalanced dataset means the percentage of the attack traffic compared to the normal traffic in the whole dataset is very low. In the following subsection, we will talk about this challenge in more details.

### 4.1 Imbalanced Dataset

Machine learning techniques like other artificial intelligence classifiers generally perform best on balanced datasets. The problem of imbalanced datasets, specifically those in severe cases (i.e., significantly low number of samples from one class compared to the other), is a critical issue in the training process. Examples of such cases include detecting rare anomalies like fraudulent bank transactions and identification of rare diseases.

Intrusion detection, which is the main security concern of IIoT applications, is another case that suffers from the severely imbalanced dataset. Due to the large amount of sensed data from IIoT devices (i.e., a large amount of normal traffic) on the one hand; and random, rare attack traffic (i.e., a small amount of attack traffic) on the other hand, the IIoT's security suffers greatly from imbalance problem.

There have been countermeasures suggested for this problem, through changing the sampling method. Under-sampling, over-sampling, or a combination of both are some examples. However, each of these techniques comes with several drawbacks. In simple terms, under-sampling means including fewer instances from the majority class, and over-sampling means including more samples of the minority class. One problem with under-sampling is the possibility of losing useful information, while over-sampling might cause overfitting problems. These techniques can be very complex, and these details are out of the scope of this paper.

Since, in a real IIoT system, any of these techniques might lead to an unrepresentative model, the resulting models may not be accurate to solve the intrusion detection problem in practice. Plus, they result in different outcomes compared to the models trained with the full dataset.

There are other challenges that must be considered when training ML techniques for intrusion detection. Here, we briefly mentioned the ones that are most critical for IIoT applications. Due to all these reasons, the suitability of ML under different circumstances must be considered. In the next section, we study the limits on the imbalanced dataset challenge on our built IIoT testbed to show the real restrictions, when it comes to training ML-based IDSs.

## 5 Our Experimental Design

In this section, we describe our testbed and the designed ML-based IDS system to show the efficiency of ML algorithms on an imbalanced IIoT dataset. In this system, intrusion detection is being evaluated by monitoring the system's transactions to detect manipulated commands.

### 5.1 Our Real-World Testbed Implementation

Utilization of a real testbed allows conducting real cyber-attacks and collecting a real dataset containing both normal and attack traffic. Considering the primary function of an ICS is to provide remote monitoring and automated control of industrial processes, we have emulated a real-world IIoT control system. Fig. 1 shows the platform of our testbed.

We chose a popular IIoT system that supervises the water level and turbidity quantity in the water storage tank. This type of system is employed in industrial reservoirs and water distribution as a part of the water treatment and distribution process. This testbed includes components like historian logs, human-machine interference (HMI), programmable logic controllers (PLCs), a three-light alarm, sensors (e.g., water levels and turbidity), actuators (e.g., alarms, valve, pumps, and buttons), and control buttons (On, Off, Light Indicator).

The main purpose of HMI in an ICS is to make it easy for the operators to observe the status of the system, interact with the IIoT devices, and receive alarms indicating abnormal behaviors. Moreover, since the sensors and relays cannot communicate directly, PLCs are used to collect the sensed data and send commands to the actuators.

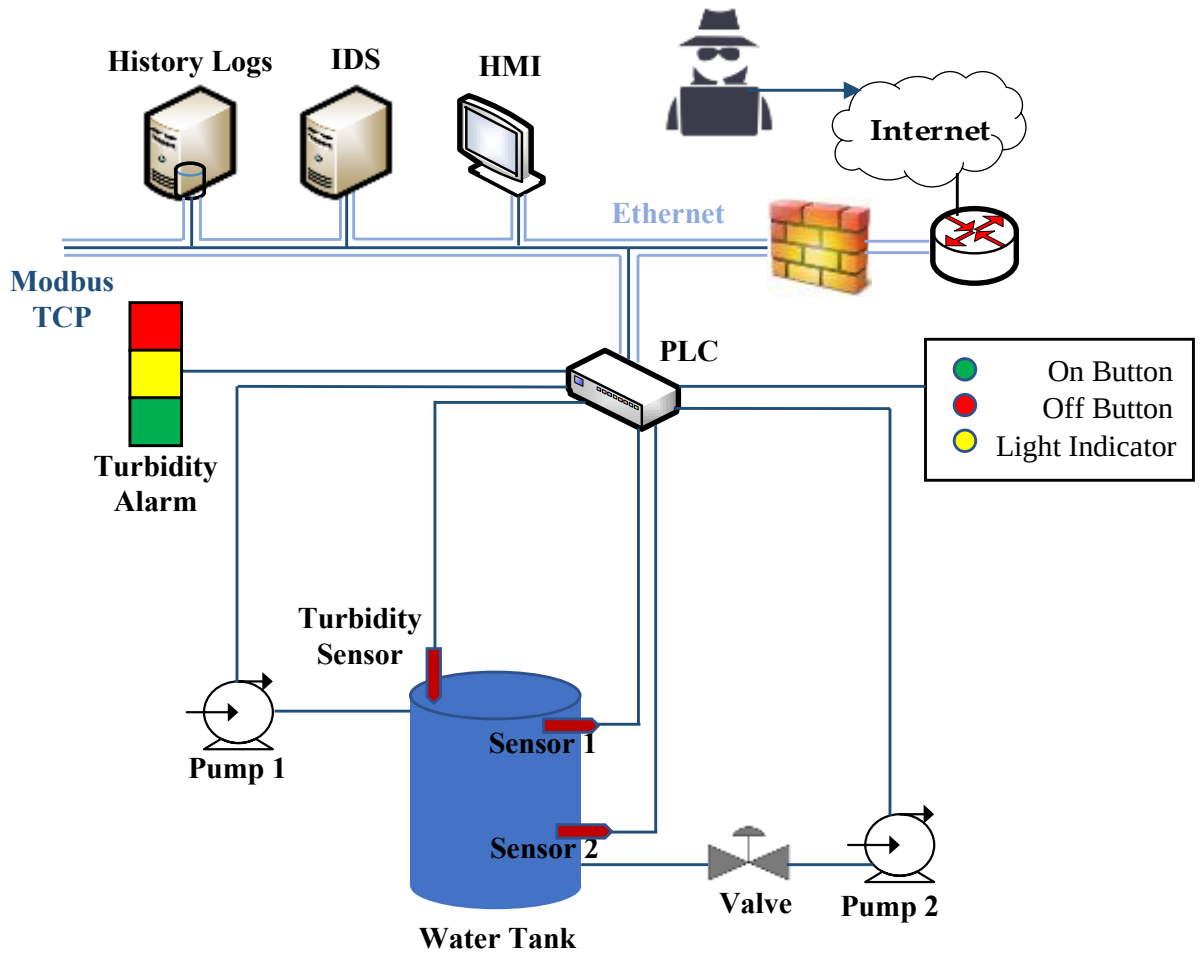

Fig. 1. Scheme of our implemented testbed

The water storage tank has two level sensors: Sensor 1 and Sensor 2, which are used to monitor the water level in the tank. When the water reaches the maximum defined level in the system, Sensor 1 sends a signal to the PLC. The PLC turns off the water Pump 1 that is used to fill up the tank, opens the valve, and turns on the water Pump 2 draws water from the tank. When the water reaches the minimum defined level in the system, Sensor 2 sends a signal to the PLC. PLC closes the valve, turns off the Pump 2, and turns on the Pump 1 to fill up the tank. This process starts over again when the water level reaches the maximum level. Meanwhile, there is an analog turbidity sensor that is integrated into the system to measure the turbidity of the water. Based on two defined thresholds in the system, PLC illuminates one of the red, yellow or green lights of the Turbidity Alarm, in which green means the water has an acceptable level of turbidity, red means the turbidity is beyond the acceptable thresholds, and yellow falls between the two thresholds.

This IIoT testbed takes the data from sensors, and the status of the system from the PLC using the Modbus communication protocol and displays them to the operator through the HMI interface. Since Modbus is one of the most popular IIoT protocols, and it's widely used by large industries, we chose this protocol.

The PLC model used in our testbed is Schneider Electric Programmable Logic Controller model M241CE40. The analog expansion module is TM3AM6 Modicon I/O Module. The logic of the PLC is programmed using the Ladder language [13], [14]. The turbidity sensor is SEN0189, and the water level sensors are Autonics CR18-8DP sensors. The deployed water pumps are GA-2328ZZ uxcell pumps.

### *5.2 Our Attack Scenario*

In this research work, we focused on the command manipulation attack in a water storage scenario to compromise the output commands. These attacks were carried out using the Kali Linux Penetration Testing Distribution using special programs for malicious command injection in ICSs. All data generated during the attacks as well as regular traffic (without attacks) was gathered and recorded by Argus [15] and Wireshark [16] network tools.

During the command injection attack, our target is the PLC. First, the attacker connects to the network to be able to read all the PLC register values and log them into a .txt file. After having the PLC register information, the attacker rewrites some of the PLC registers that are vital to the physical process. For example, we ran this attack while Pump 2 was supposed to draw water from the tank, it was suddenly stopped by the attacker, and Pump 1 started, and the water overflowed from the tank. Another instance is when the attacker turned on the wrong turbidity alarm light, in the way that, while the turbidity level was high, and the red light was supposed to be on, the attacker turned off the red light and turned on the green light instead.

### *5.3 Feature Selection*

In this case study, we use Artificial Neural Network (ANN). The model is trained and tested over the imbalanced dataset collected from our testbed, and the results of their performance are compared (details in the next Subsection).

TABLE I. SELECTED TRAFFIC FEATURES IN OUR PROPOSED IDS

| Features | Type | Descriptions |
|---|---|---|
| Mean flow (mean) | Float | The average duration of active flows |
| Source Port (Sport) | Integer | Source port number |
| Destination Port (Dport) | Integer | Destination port number |
| Source Packets (Spkts) | Integer | Source/Destination packet count |
| Destination Packets (Dpkts) | Integer | Destination/Source packet count |
| Total Packets (Tpkts) | Integer | Total transaction packet count |
| Source Bytes (Sbytes) | Integer | Source/Destination bytes count |
| Destination Bytes (Dbytes) | Integer | Destination/Source bytes count |
| Total Bytes (TBytes) | Integer | Total transaction bytes count |
| Source Load (Sload) | Float | Source bits per second |
| Destination Load (Dload) | Float | Destination bits per second |
| Total Load (Tload) | Float | Total bits per second |
| Source Rate (Srate) | Float | Source packets per second |
| Destination Rate (Drate) | Float | Destination packets per second |
| Total Rate (Trate) | Float | Total packets per second |
| Source Loss (Sloss) | Float | Source packets retransmitted/dropped |
| Destination Loss (Dloss) | Float | Destination packets retransmitted/dropped |
| Total Loss (Tloss) | Float | Total packets retransmitted/dropped |
| Total Percent Loss (Ploss) | Float | Percent packets retransmitted/dropped |
| Source Jitter (ScrJitter) | Float | Source jitter in millisecond |
| Destination Jitter (DrcJitter) | Float | Destination jitter in millisecond |
| Source Interpacket (SIntPkt) | Float | Source interpacket arrival time in millisecond |
| Destination Interpacket (DIntPkt) | Float | Destination interpacket arrival time in millisecond |

An important step in training the algorithm is selecting and extracting features from the raw network traffic traces. Here, in designing our IDS, we chose 23 features. These features are common in network flows and also show a good variation during the attack phases. Table I shows the chosen features along with their description.

How each feature varies depends on the type of the attack. For instance, during the normal condition, where no attack is conducted, the SrcPkts and DstPkts features mostly show a periodic behavior. Meanwhile, during attacks, these features behave randomly.

### *5.4 Imbalance Setting*

Our main goal here is to examine the efficiency of ANN in detecting anomaly through different imbalance ratios. We collected a new dataset of 2.7 GB, for a total of about 53 hours. The number of attacks at each trial has been kept equal to 10000 samples, and accordingly, we added normal traffic to build the desired ratios. Table II is a summary of the number of samples used. At each round of training, we divided the dataset into 80% for training and 20% for testing.

TABLE II. OUR BUILT DATASET STATISTICAL INFORMATION

| Ratio | # of Attack | #of Normal | Total |
|---|---|---|---|
| 10.0% | 10,000 | 90,000 | 100,000 |
| 1.0% | 10,000 | 990,000 | 1,000,000 |
| 0.7% | 10,000 | 1,418,572 | 1,428,572 |
| 0.3% | 10,000 | 3,323,334 | 3,333,334 |
| 0.1% | 10,000 | 9,990,000 | 10,000,000 |

### *5.5 Performance Metrics*

Traditionally, the performance of the trained algorithms is measured by metrics which are derived from the confusion matrix. Table III shows the confusion matrix.

TABLE III. CONFUSION MATRIX IN IDS CONTEXT

| | | Predicted Class | |
|---|---|---|---|
| | | ***Classified as Normal*** | ***Classified as Attack*** |
| **Actual Class** | ***Normal Data*** | True Negative (TN) | False Positive (FP) |
| | ***Attack Data*** | False Negative (FN) | True Positive (TP) |

The description of the matrix confusion parameters is as follows:

- True Negatives (TN): Represents the number of normal packets correctly classified as normal.
- True Positives (TP): Represents the number of abnormal packets (attacks) correctly classified as attacks.
- False Positive (FP): Represent the number of normal packets incorrectly classified as attacks.
- False Negative (FN): Represents the number of abnormal packets (attacks) incorrectly classified as normal packets.

According to the confusion matrix, the metrics that are used in this work to evaluate the performance of the ML algorithms are as follows:

- Accuracy: Shows the percentage of the correctly predicted samples considering the total number of predictions.

$$Accuracy= \frac{TP+TN}{TP+TN+FP+FN} \times 100 \quad (1)$$

- False Alarm Rate (FAR): Represents the percentage of the regular traffic misclassified as attacks.

$$FAR= \frac{FP}{FP+ TN} \times 100 \quad (2)$$

- Undetected Rate (UR): The fraction of the anomaly traffic (attack) misclassified as normal.

$$UR= \frac{FN}{(FN+TP)} \times 100 \quad (3)$$

- Matthews Correlation Coefficient (MCC): Measures the quality of the classification. MCC is a great metric, especially in case of imbalanced datasets, showing the correlation agreement between the observed values and the predicted values.

$$MCC= \frac{TP\times TN\text{-}FP\times FN}{\sqrt{(TP+FP)\times(TP+FN)\times(TN+FP)\times(TN+FN)}} \times 100 \quad (4)$$

- Sensitivity: Also known as the true positive rate. A sensitive algorithm helps rule out an attack situation with more confidence when the prediction is negative.

$$Sensitivity= \frac{TP}{TP+FN} \times 100 \quad (5)$$

Accuracy (1) is the most frequently used metric for assessing the performance of learning models in regression problems. However, this metric is not sufficient for performance evaluation in scenarios with imbalanced classes (i.e., one class is dominant and has more training data compared to the other). In our case, which is an IDS scenario, the proportion of normal traffic to attack traffic is very high resembling a realistic dataset. Therefore, in addition to the accuracy, we use other metrics that represent the performance better and in a more delicate way.

### *5.6 Results*

In this section, we present the numerical results of our algorithms detecting the command injection attacks through different ratio of imbalance as it was mentioned in the previous subsection. Through all these figures, each point on the graph is marked with the corresponding ratio (e.g., “10%” means the ratio of attack samples to the normal samples is 1 to 9).

As shown in Fig. 2, representing the accuracy results (1), it seems there is not much difference in accuracy performance. However, this is not true. In intrusion detection scenarios with the imbalanced dataset, accuracy is not the best representative metric to evaluate the performance. Since a large portion of training data is normal traffic, the algorithms are biased toward estimating all the data as normal and ignoring the small portion of the attack instances.

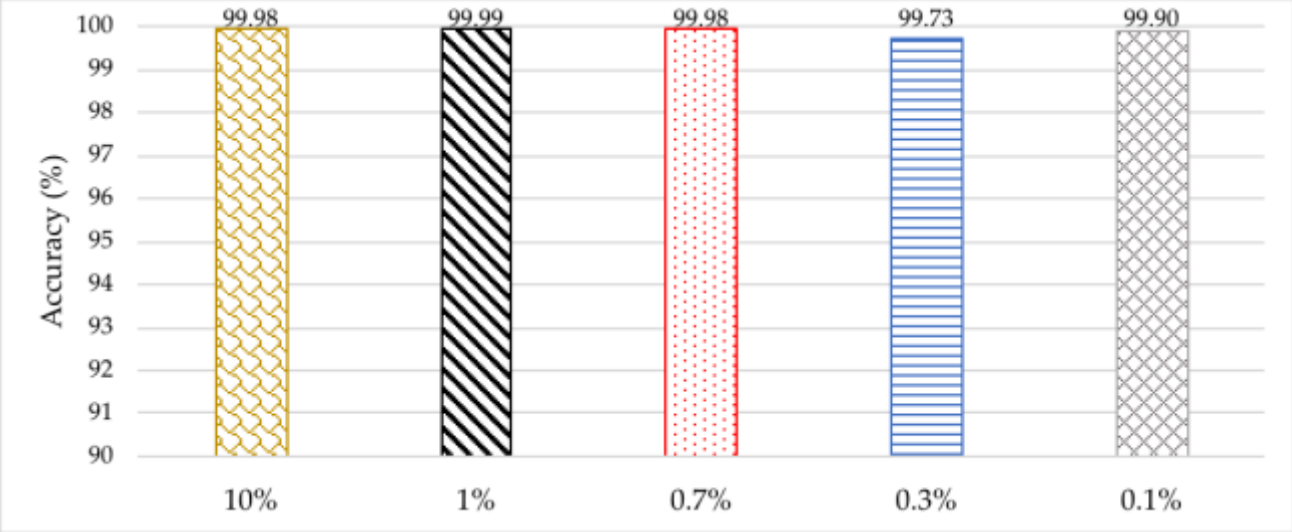

Fig. 2. Effect of Imbalance on Accuracy

The false alarm rate (FAR), as shown in Fig. 3, represents the percentage of the normal traffic being misclassified as the attack traffic by the model (2). As again it is seen, Fig. 3 shows good performance for all the cases. However, for the same reason, even this metric cannot truly represent the performance. For example, Since the amount of attack traffic is considerably low in the 0.1 % scenario, the algorithms would barely label any instances as an attack; hence, we would expect a low FAR percentage.

Undetected rate (UR) metric can assess the performance better despite being imbalanced. As shown in Fig. 4, UR represents the percentage of the traffic which is attack traffic but is misclassified as normal (the opposite of the FAR) (3). Since this metric considers only the attack traffic, the fact of having an imbalanced dataset does not impact the evaluation that much. The training with 0.1% attack training data was barely able to detect any anomaly and showed the worst performance as it was expected. This metric is more critical than FAR because it is related to the attacks that happen without being detected by the system.

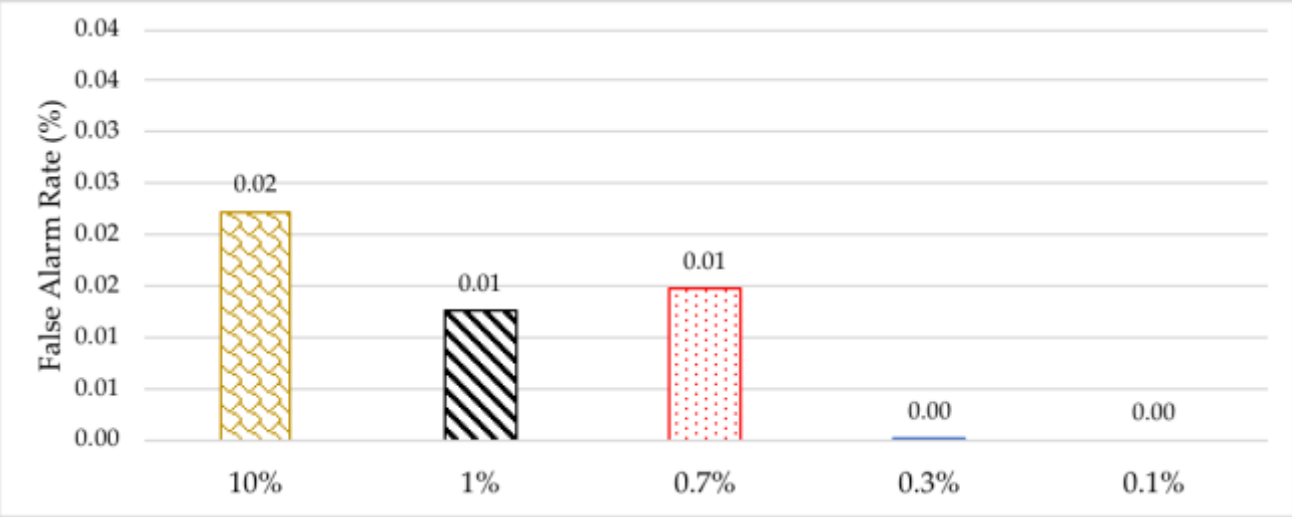

Fig. 3. Effect of Imbalance on False alarm rate

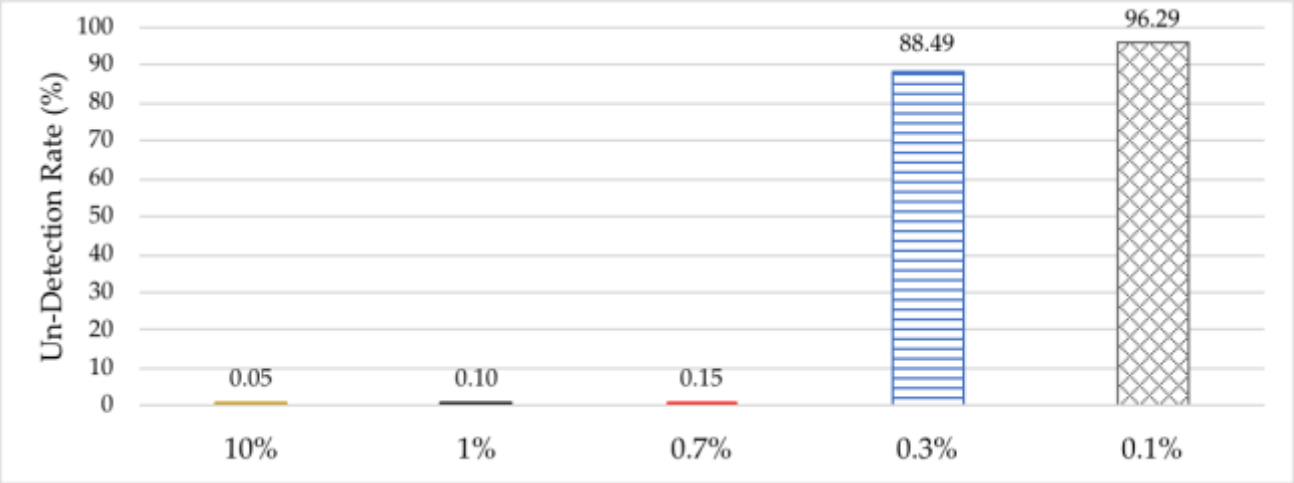

Fig. 4. Effect of Imbalance on Undetected Rate

MCC (5) is considered to be one of the best metrics for classification evaluation, and it is generally a better performance representative compared to the other metrics. As shown in Fig. 5, the more attack data is used for training, the better the MCC value we would get. MCC is considered as an appropriate metric when it comes to evaluating ML models that are trained with an imbalanced dataset.

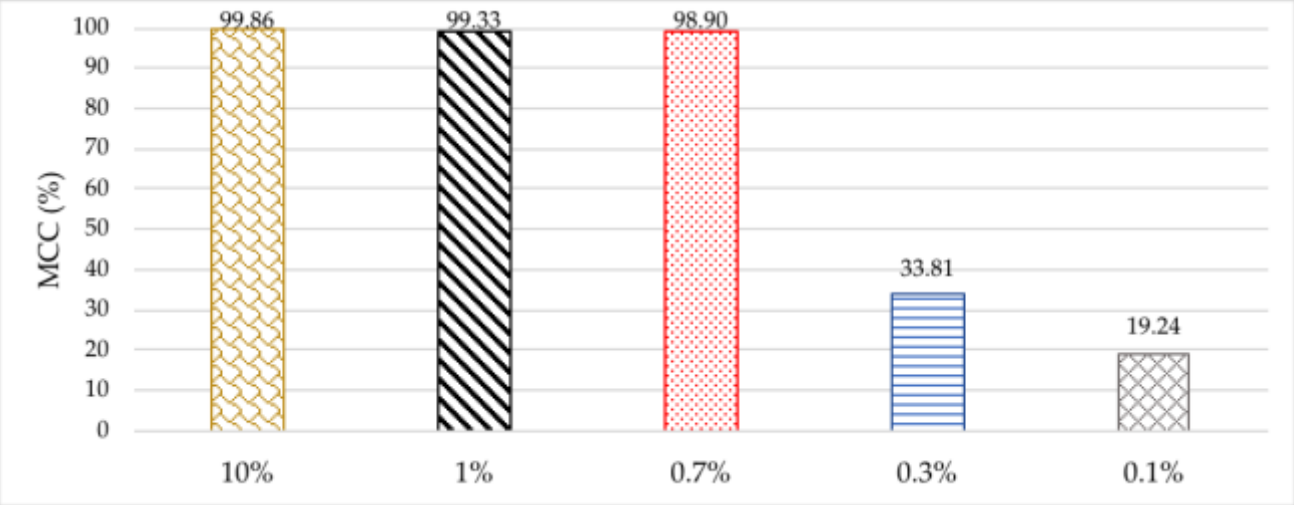

Fig. 5. Effect of Imbalance on MCC

Finally, the sensitivity metric results (6) are shown in Fig. 6 to evaluate how sensitive the model is to able to react to an abnormal situation. As seen in the figure, training with more abnormal traffic will result in showing more sensitivity in the detection performance.

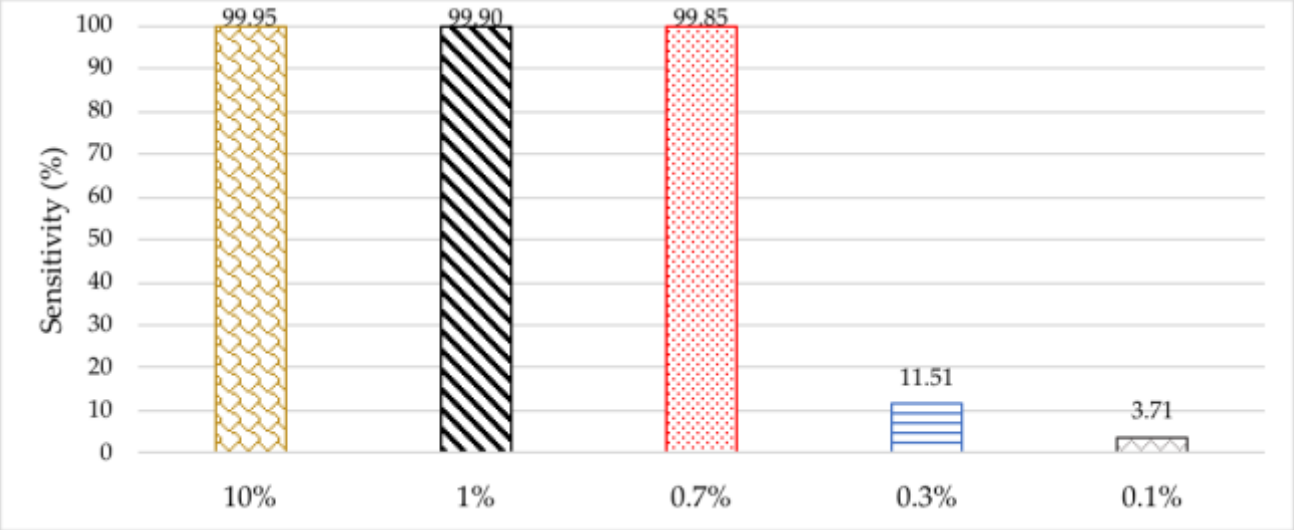

Fig. 6. Effect of Imbalance on Sensitivity

To provide a performance comparison with a baseline sampling method, we chose synthetic minority over-sampling technique (SMOTE). In this technique, we synthesize new fake attack data from the existing attack samples based on their *k* nearest neighbors. For more information, we refer the readers to [17]. We ran this method only for 7%, 3%, and 1% anomaly ratios since these three were the severe cases.

Fig. 7 shows the undetected rate before and after using the SMOTE technique. As it is shown in this picture, this method decreased the rate of undetected attacks to 0 for 0.7% and 0.3% imbalance ratios and to about 57% in the 0.1% case. Even though through this technique at 0.1%, we achieved a better rate, still more than half of the attack data were not discovered.

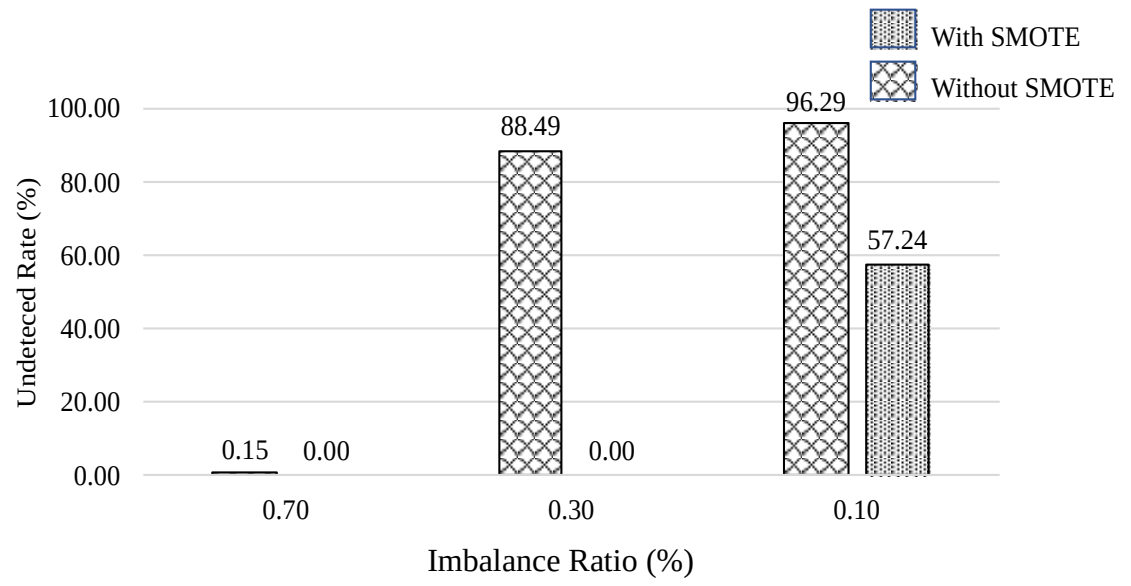

Fig. 7. Effect of SMOTE Method on Undetected Rate

The MCC results are shown in Fig. 8. We observe a great improvement in the 0.3% case, with a slight degradation with the 0.7% imbalance ratio. As a result, this oversampling technique helped the system distinguish the attack scenarios more effectively and perform better in low imbalance ratios.

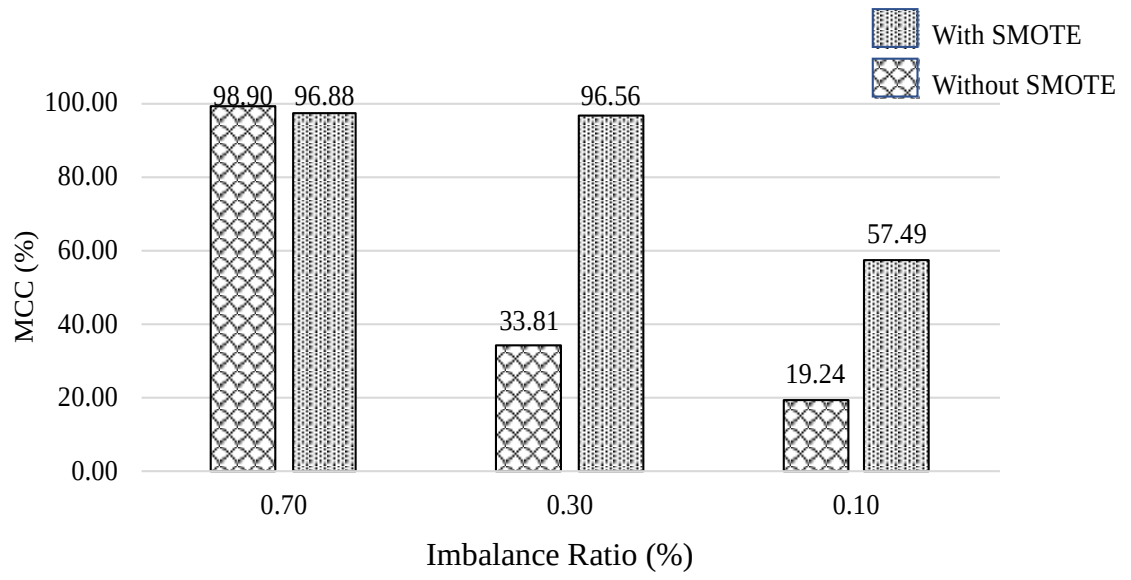

Fig. 8. Effect of SMOTE Method on MCC

## 6 Conclusion

The cyber-security of the IIoT devices is critical. Intrusion detection is the main security concern in these applications. Machine learning solutions and big data analytics have been widely used to ensure a secure platform in these systems. However, when it comes to a real-world scenario and applying these algorithms practically, they sometimes fall short. The main focus of this paper was studying imbalanced dataset problems and show in which extend the machine learning algorithms are able to help.

## Acknowledgment

This work has been supported under the grant ID NPRP 10-901-2-370 funded by the Qatar National Research Fund (QNRF) and grant#2017/01055-4 São Paulo Research Foundation (FAPESP). The statements made herein are solely the responsibility of the authors.